\documentclass[preprint,showpacs,preprintnumbers,amsmath,amssymb]{revtex4}

\usepackage{graphicx}
\usepackage{dcolumn}
\usepackage{bm}

\begin{document}

\preprint{Nikolay K. Vitanov}

\title{Analysis of a Japan government intervention on the 
domestic agriculture  market}

\author{Nikolay K. Vitanov}
\affiliation{Institute of Mechanics, Bulgarian Academy of Sciences, Akad. G. Bonchev
Str., Bl. 4, 1113 Sofia, Bulgaria}
\email{vitanov@imbm.bas.bg}
\author{Kenshi Sakai}%
\affiliation{Department of Eco-Regional Science, Tokyo University of Agriculture and Technology,
3-5-8, Saiwai-cho, Fuchu-shi, Tokyo, 183-8509, Japan}
\author{Ivan P. Jordanov}
\affiliation{LPCH, Institute of Mechanics, Bulgarian Academy of Sciences, Akad. G. Bonchev
Str., Bl. 4, 1113 Sofia, Bulgaria}
\author{Shunsuke Managi}
\affiliation{Faculty of Business Administration, Yokohama National University,
79-4, Tokiwadai, Hodogaya-ku, Yokohama, 240-8501, Japan}
\author{Katsuhiko Demura}
\affiliation{Department of Agricultural Economics, Hokkaido University, N9, W9,
Sapporo 060, Japan}

\date{\today}

\begin{abstract}
We investigate an economic system in which one large agent - the Japan government
changes the environment of numerous smaller agents - the Japan agriculture producers
by indirect regulation of prices of agriculture goods. The reason
for this intervention was that 
before the oil crisis in 1974 Japan agriculture production prices 
exhibited irregular and large amplitude changes.  
By means of analysis of correlations and  a combination of
 singular spectrum analysis (SSA),
principal component analysis (PCA), and time delay phase space construction
(TDPSC) we study the influence of the
government measures on the domestic piglet prices and production 
in Japan.  We show that the government
regulation politics was successful and leaded 
(i) to a decrease of  the nonstationarities and to increase of predictability
of the piglet price;  
(ii)  to a coupling of the price and production cycles;  
(iii) to increase of determinism of the dynamics of the fluctuations
of piglet  price around the year average price. The investigated
case is an example confirming the thesis that a large agent can
change in a significant way the environment of the small agents in complex (economic
or financial) systems which can be crucial for their
survival or extinction.
\end{abstract}

\maketitle

\section{\label{sec:level1} Market complexity and government regulation}
The growing complexity of the human society increases the difficulties 
in ensuring its steady economic development. This leads to
increasing importance of large social agents such as state central
bank or the national government which can influence in a significant
way the financial and economic subsystems of the society. Changes of the 
interest rates, intervention on the exchange markets and subsidies for
selected economic branches improve or worsen 
the environment for the small size economics
and social agents. Below we discuss an especially critical issue for 
the sustainable development of the society namely the 
development of the large network of economic, ecological and social 
components called national agricultural system. A sudden decrease 
of the agriculture production can lead to social tensions. The requirement for steady increase of the food quality and quantity 
presses entire branches of the  agriculture system to move from the 
natural (due to the climate) regimes of cyclic output to more uncontrollable 
regimes of chaotic output. Because of the above two reasons the
governments try to regulate the market for agriculture goods especially 
after large crises such as wars or oil shocks 
\cite{cohrane,spulber,colyer}. Here we shall analyze the results of 
a such intervention by the methods of the time series analysis
(see \cite {kantz} and for  applications 
\cite{d1,d2,boeck}). Together with the methods of the nonlinear 
dynamics and stochastic analysis the above-mentioned methods 
are much applied  for understanding, description and evaluation
of the dynamics of complex economic and social systems 
\cite{lorenz,levy,v1,v2} and in particular 
in studies of different aspects of the agriculture production such as
analysis of the dairy production in the USA \cite {chavas}
or the analysis of the meat production in Japan \cite{sakai}. 
\par
With its about 130 million citizens and highly developed economy
Japan is a large consumer of meat but up to about 45 years ago 
no specialized in piglet  production farms  existed. As the
consumption increased in the 1960's  this production
became lucrative source of money for the farmers and
specialized farms for production of piglets and meat arose. 
Until 1974 there was no
massive intervention of the Japanese government on the agriculture market, i.e.,
less than $1/10$ from the farmers received subsides from the state. After
the oil crisis from 1974 the situation changed significantly and almost all
agricultural producers have been included in the subsidies schemes
and other farm programs intending to fix the instability of the agriculture markets.
\section{Was the Japan government intervention successful?}
In order to investigate the consequences of the actions
of the large agents often we do not need to analyze the
large amount of data normally required by econophysicsists
for analysis of systems containing many almost identical 
small-size agents. Here we study
monthly time series for the piglet production and prices in Japan
from 1965 to 1992 (i.e. before the recession
events in the last decade of previous century). The time series are too short and 
too nonstationary in order to be investigated by means of the
conventional methods of the nonlinear time series analysis such as
Lyapunov exponents or generalized dimensions but we can apply 
another methodology based on singular spectrum analysis (SSA) and 
principal component analysis (PCA) (see for
example \cite{broom,albano,mees,vautard,v3}). 
\begin{figure}
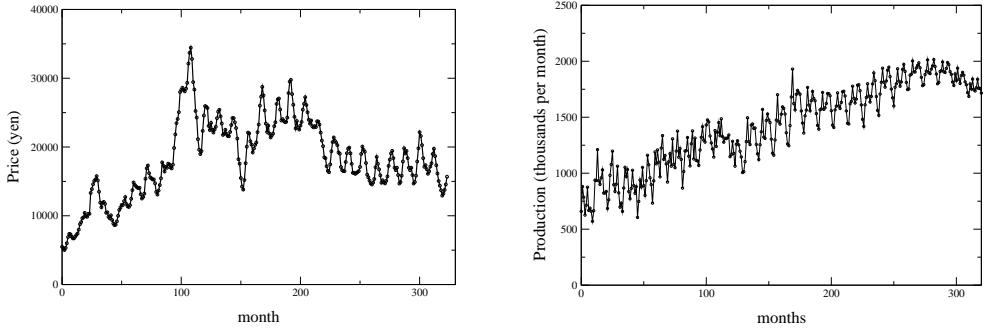

\vskip1cm
\begin{center}
\includegraphics[width=6cm]{price.eps}\hskip1cm
\includegraphics[width=6cm]{production.eps}
\end{center}
\vskip1cm
\caption{Piglet prices and production in Japan, 1965-1992.}
\end{figure} 
The time series are shown in Fig.1. The oil crisis
from 1974 is located just before the maximum of the price time series. In the 
period before the crisis the price time series are very 
nonstationary with a sharp upward trend. The
government intervention after the crisis stabilized the price and
the seasonal periodicity became more important in comparison to the period
before the oil crisis. Such drastic changes
are not observed in the time series for the production which shows a trend of increase up
to the beginning of 1990s.
\begin{figure}
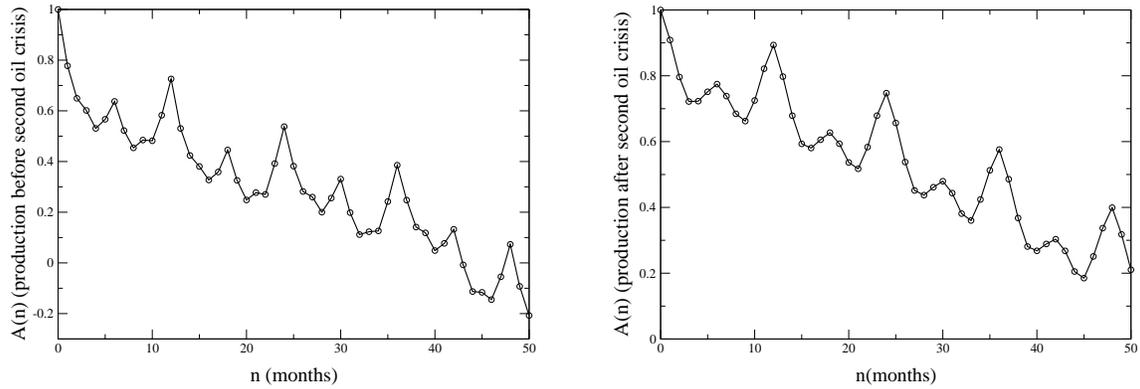

\vskip1cm
\begin{center}
\includegraphics[width=7cm]{acor_production_before_crisis.eps}\hskip1cm
\includegraphics[width=7cm]{acor_production_after_crisis.eps}
\end{center}
\caption{Autocorrelation functions for the piglet production before 
(left-hand side) and after (right-hand side) the oil crisis in 1974.}
\vskip1cm
\end{figure}
The autocorrelation function supplies us with additional information. 
For an example  if the time series are correlated and not periodic 
the autocorrelation function decays slower than exponentially 
and in presence of long-range correlations we can observe a 
power-law decay (for more details see for an example 
\cite{bass,s1,s2,v4,v5}).  
\begin{figure}
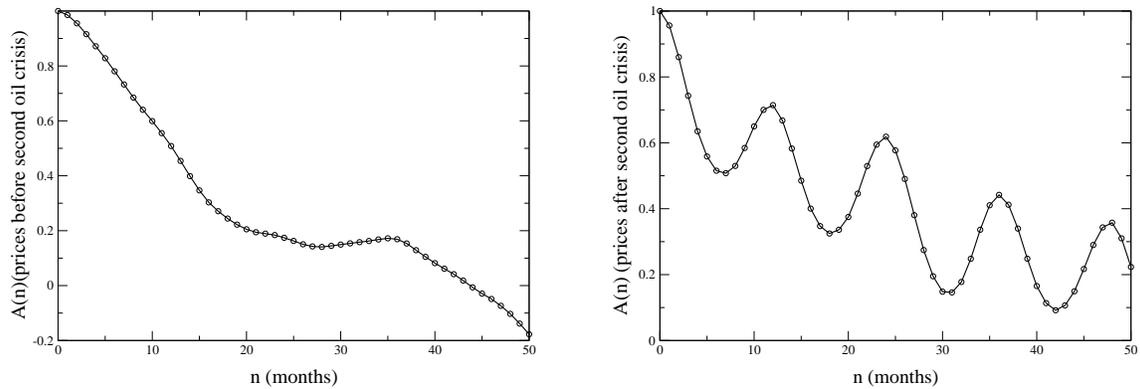

\vskip1cm
\begin{center}
\includegraphics[width=7cm]{acor_prices_before_crisis.eps}\hskip1cm
\includegraphics[width=7cm]{acor_prices_after_crisis.eps}
\end{center}
\caption{Autocorrelation functions for the piglet prices before (left-hand side panel) 
and after (right-hand side panel) the oil crisis in 1974.}
\vskip1cm
\end{figure}
The autocorrelation functions for the production and prices
of the piglets in Japan are shown in Figs. 2 and 3.  In Fig. 2 we observe 
slower decay of the autocorrelation function $A(n)$ of the production time series
after the oil crisis when compared to the pre-crisis function. 
Thus after the government intervention
the piglets production  became more correlated in the time, e.g.,
more predictable. The government intervention has not influenced significantly
the (caused by the specifics of the consumers demand)  6- and 12-months
production cycles. 
Completely different is the situation about the piglet prices. Before 
the crisis the price correlations decay with increasing  $n$. 
After the government intervention the strong upward trend 
is not presented anymore. 
The dominant process is  a 12-months cycle and the 
minima and maxima of $A(n)$ follow each
other in a 6-months tact. Thus the intervention coupled the price and production cycles
and the prices became much more predictable.
\par
 In combination with the  method of delay vector construction \cite{kantz, hegger} the singular spectrum analysis can deal successfully with short and
nonstationary time series \cite{broom,albano,mees,vautard,v3}. 
Let us have a time series consisting of $N^{*}$ values $x(\tau_{0}),x(2\tau_{0}),
\dots, x(N^{*} \tau)$ recorded by using  fixed time step $\tau_{0}$. On the basis of the
time series we construct $m-$dimensional vectors as follows. First we choose the step
$\tau=n \tau_{0}$ and then we construct the vectors  
$\vec{X}_{i}=\{ x(i \tau_{0}),x(i \tau_{0} + \tau), \dots, x(i \tau_{0} + 
(m-1)\tau)\}$. In such a way by means of the TDPSC (time delay phase 
space construction) we transform the time series to a set of vectors. We note 
that the requirements for TDPSC are different from the requirements for the
time-delay embedding. In TDPSC we have to choose $n$ as small as possible
and $m$ must be as large as possible (here $n=1$, $m=30$). 
By means of the TDPSC vectors we build the trajectory matrix
${\bf X} = \frac{1}{N^{1/2}}[\vec{X}_{1}^{T},\vec{X}_{2}^{T},\dots,\vec{X}_{N}^{T}]^{T}$
as well as the covariance matrix of the trajectory ${\bf K} = {\bf X}^{T} {\bf X}$.
Let $\vec{k}_{p}$ be the eigenvectors of the last matrix and $\sigma_{p}$ are the
eigenvalues corresponding to these vectors. The vectors $\vec{k}_{p}$ form an orthonormal
basis in the $m-$ dimensional space of the vectors $\vec{X}_{i}$. The matrix ${\bf X}$
can be decomposed in the following way: ${\bf X} = {\bf S} {\bf \Sigma} {\bf C}^{T}$
where ${\bf S}$ is an $N \times m$ matrix consisting of the eigenvectors of the trajectory
matrix \cite{broom}. ${\rm {\bf C}}=[\vec{k}_{1}, \vec{k}_{2}, \dots,
\vec{k}_{m}]$
is the $m \times m$ orthogonal matrix and ${\bf \Sigma} = {\rm diag[\sigma_{1},\sigma_{2},
\dots, \sigma_{m}]}$ is the diagonal matrix constructed by the eigenvalues $\sigma_{i}$
called also singular values. These values are nonnegative and the common rule is to number them
with respect to their values as follows: $\sigma_{1} \ge \sigma_{2} \ge \dots \ge 
\sigma_{m} \ge 0$.
In addition we can decompose the time series $\rm \{ x_{i} \}$ using the eigenvectors $\vec{k}_{q}$ of the 
Toeplitz matrix of the time series
$x_{i+j}=\sum_{l=1}^{m} a_{i}^{l} k_{j}^{l}; 1 \le j \le m$.
The principal components $a_{i}^{l}$ of the time series can be obtained
by a projection of the time series on the basis vectors
$a_{i}^{l} = \sum_{j=1}^{m} x_{i+j} k_{j}^{l}$.
\begin{figure}[h]
\vskip1cm
\begin{center}
\includegraphics[width=7cm]{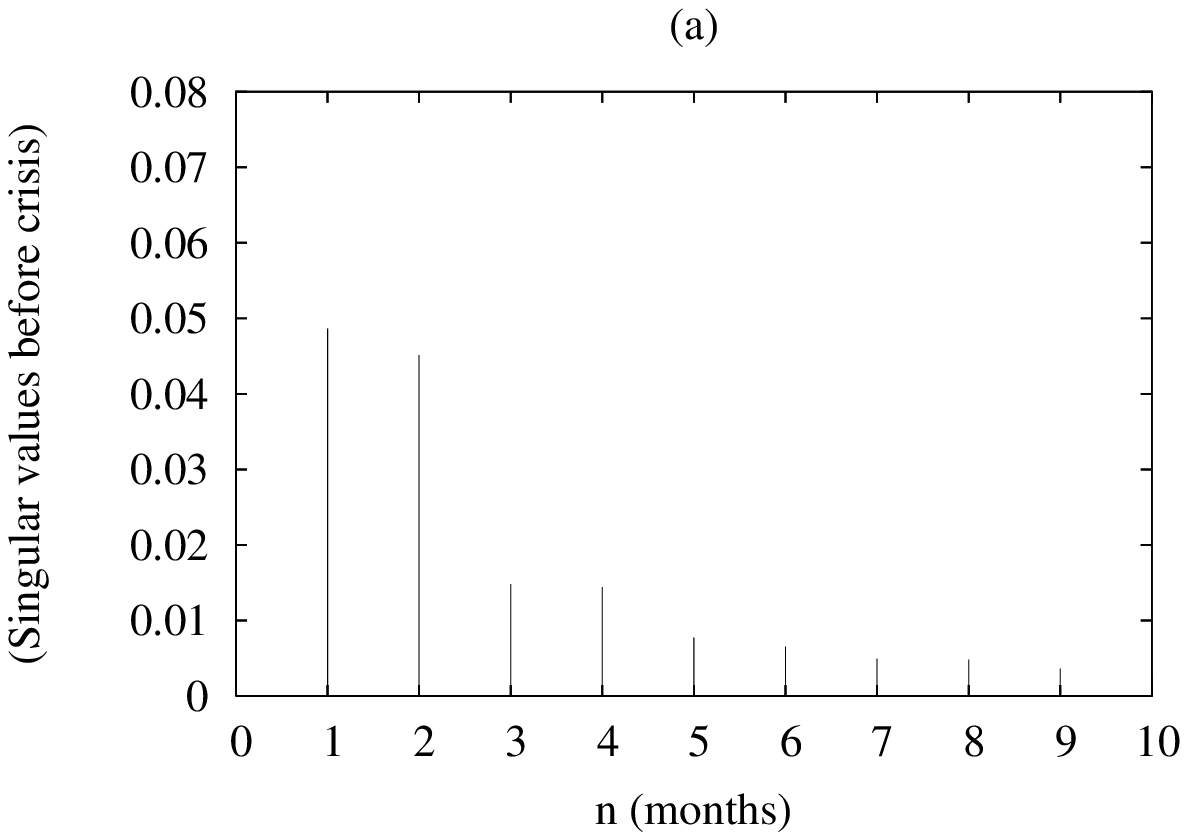}\hskip1cm
\includegraphics[width=7cm]{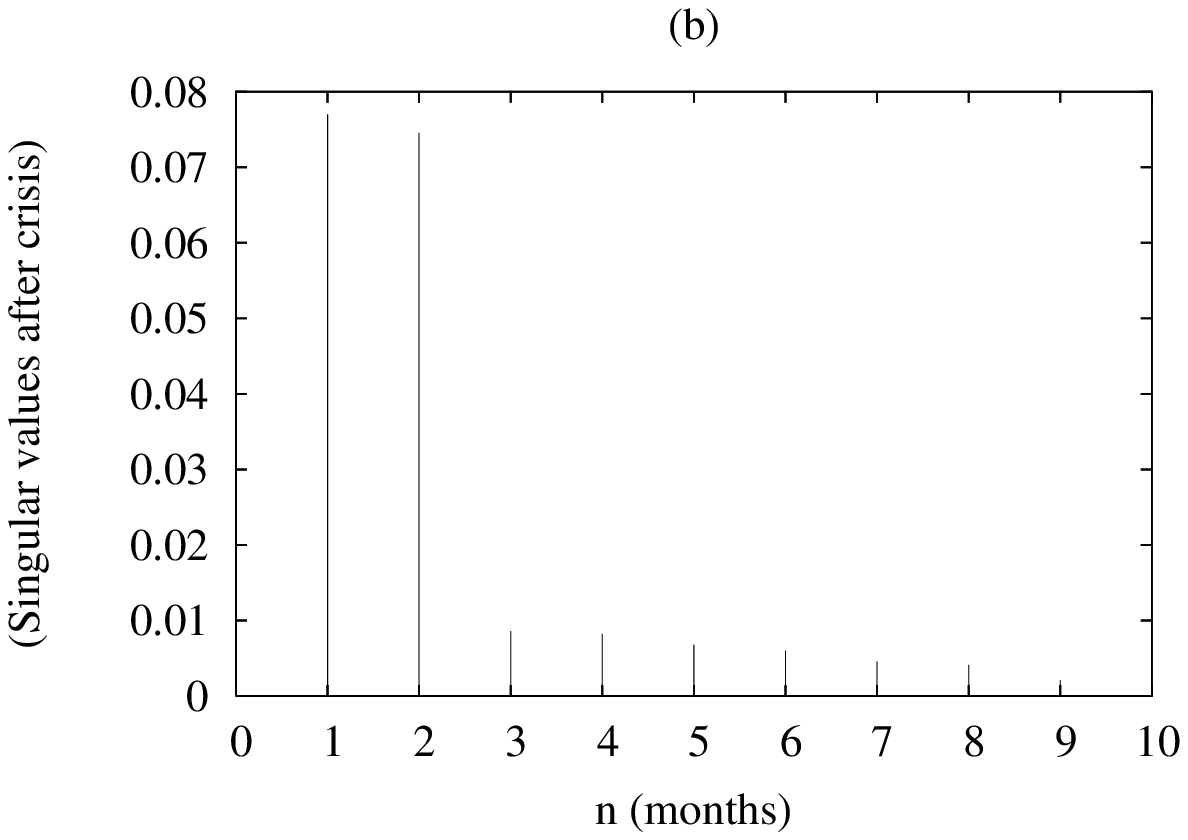}
\end{center}
\caption{Singular values for the detrended price time series before (left-hand
side panel) and after (right-hand side panel) the  crisis.}
\vskip1cm
\end{figure}
\begin{figure}[h]
\vskip1cm
\begin{center}
\includegraphics[width=7cm]{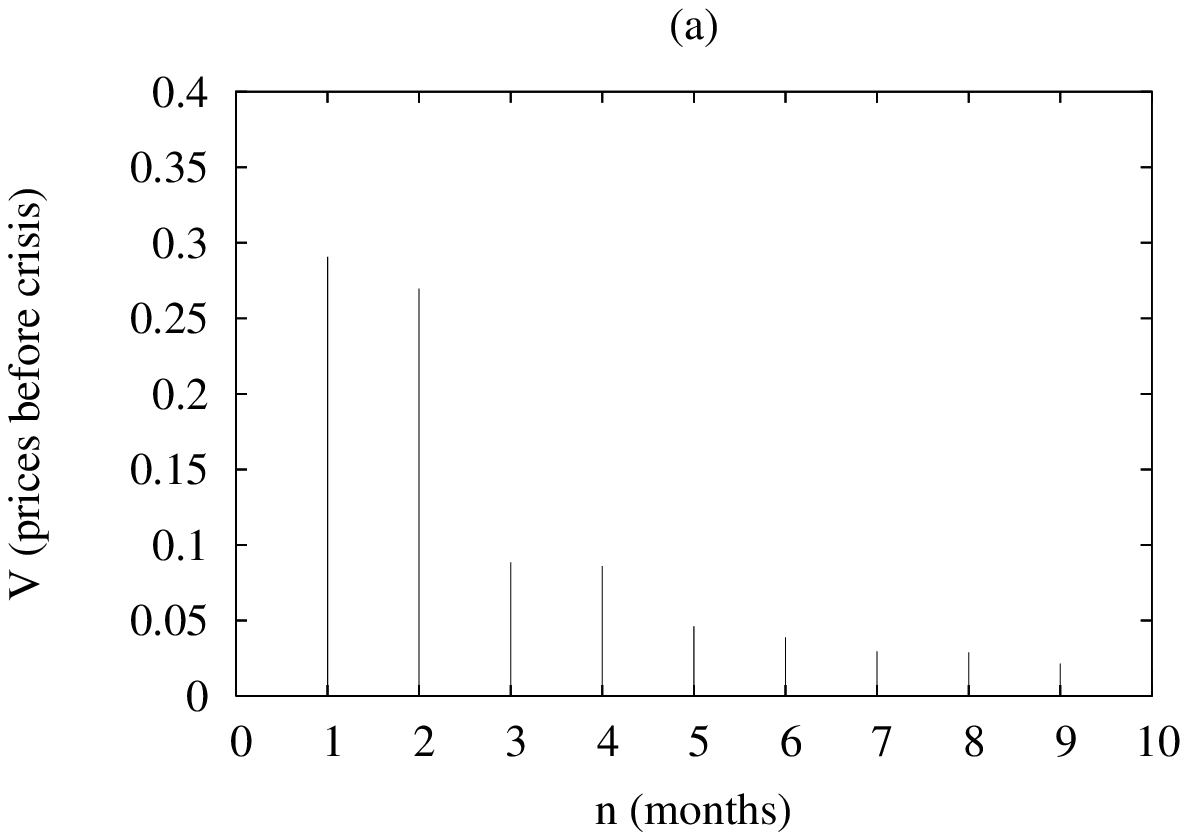} \hskip1cm
\includegraphics[width=7cm]{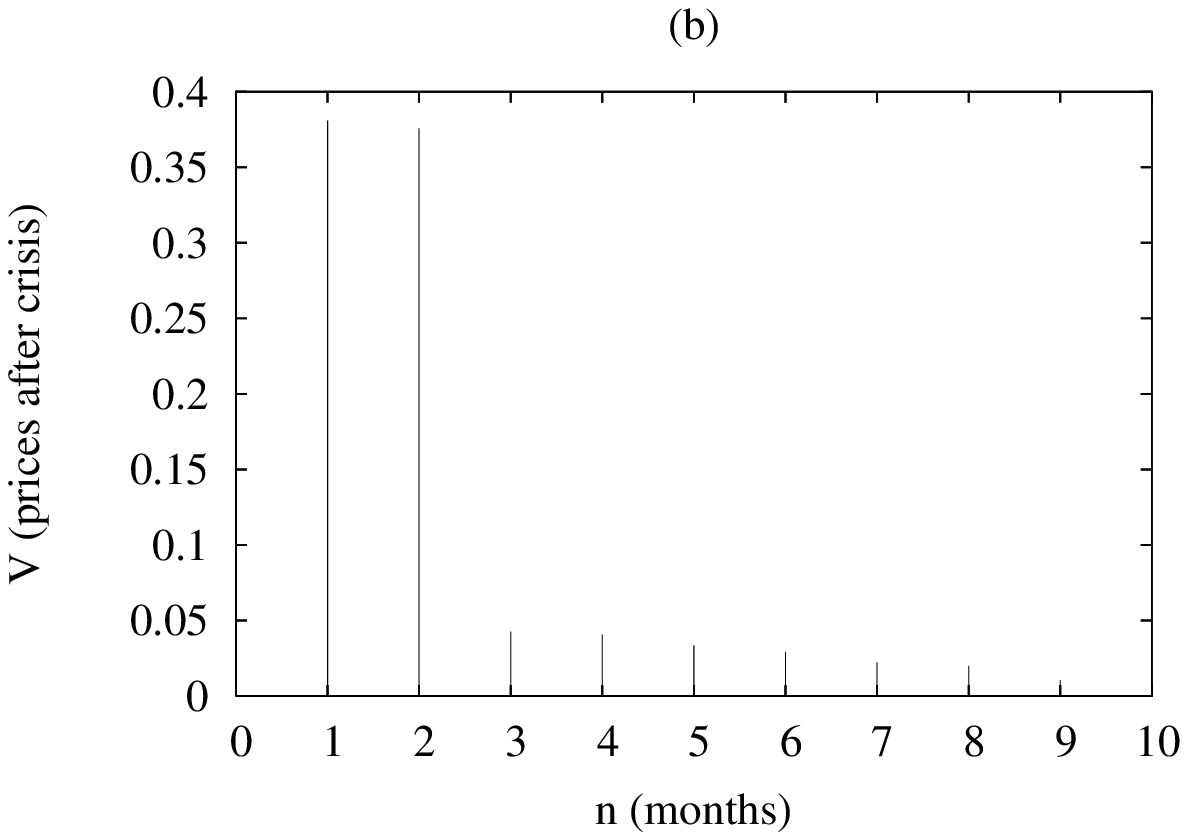}
\end{center}
\caption{Variance in the principal components for the detrended time series of the
piglet prices before and after the crisis.}
\vskip1cm
\end{figure}
\par
Below we shall extract additional information about the behavior of the 
piglet prices by application of SSA and PCA to
the piglet price time series from which the
moving average year trend is subtracted. What remains are the oscillations around the
trend and we shall investigate if the the dimension of the
phase space dynamics of these oscillations decreased as a result  of 
the discussed Japan government intervention. Strong evidence for
such a conclusion is obtained by comparing the singular spectra
of the the time series as well as the part of total variance of time series contained in the corresponding principal components. These two quantities for
the detrended time series before and after the crisis are shown in Figs. 4 and 5.
The largest singular values in the singular spectra shown in Fig. 4
are connected to the deterministic part of the time series  and the number of the
significant components in the singular spectrum (which values are
significantly larger than the other singular values) gives us the
value of the statistical dimension $S$ which is an upper bound of
significant degrees of freedom of the investigated system. 
The concentration of fluctuation dynamics can be observed in Fig.4 where we
can see the increase of the significance of the first two principal components
after the crisis at the expense of the significance of the second two principal
components. Additional light on the concentration of the dynamics is given by 
Fig. 5 where we observe increasing of the percentage of total variance of time series
concentrated in the first two principal components after the crisis.  We can conclude that the dynamics of the fluctuations
is (a) low -dimensional even before the crisis and (b) its dimensionality decreases further
after the crisis and the essential part of the dynamics can be captured by an
appropriate model containing small number of  equations. 
\begin{figure}[h]
\vskip1cm
\begin{center}
\includegraphics[width=7cm]{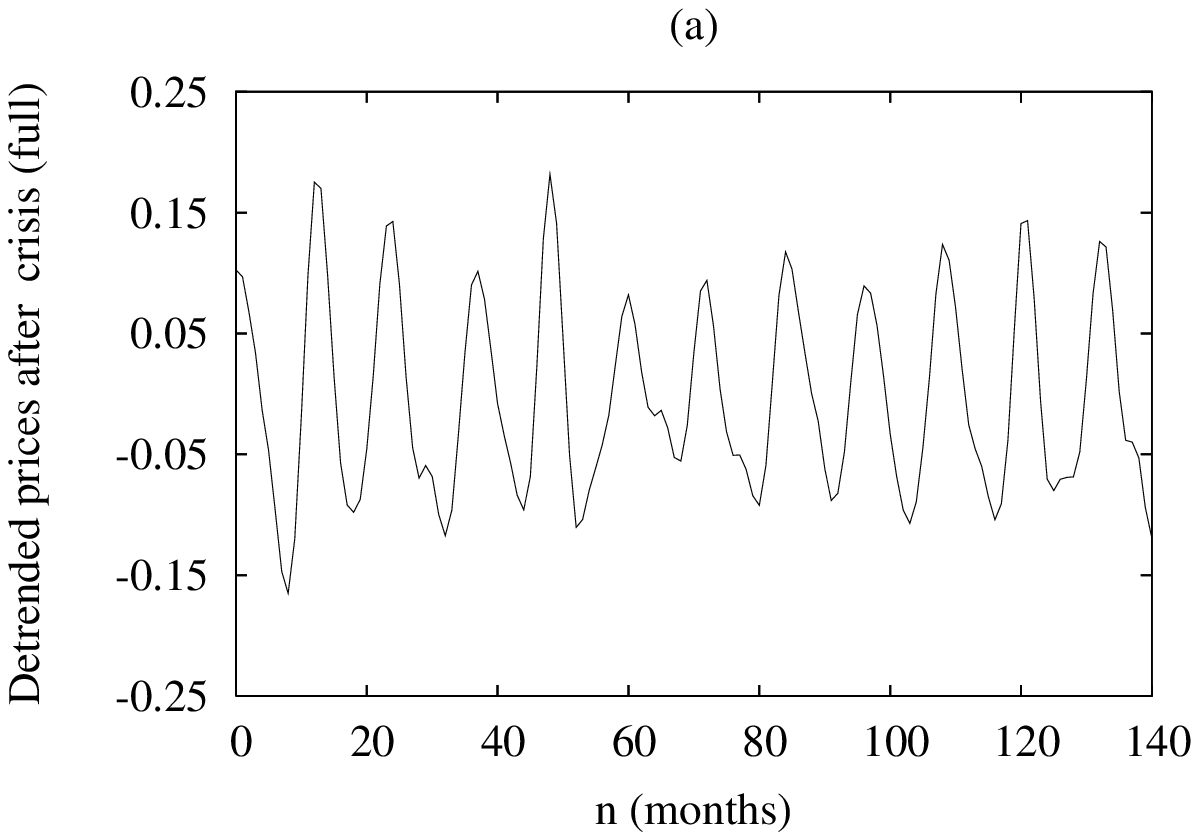}\hskip1cm
\includegraphics[width=7cm]{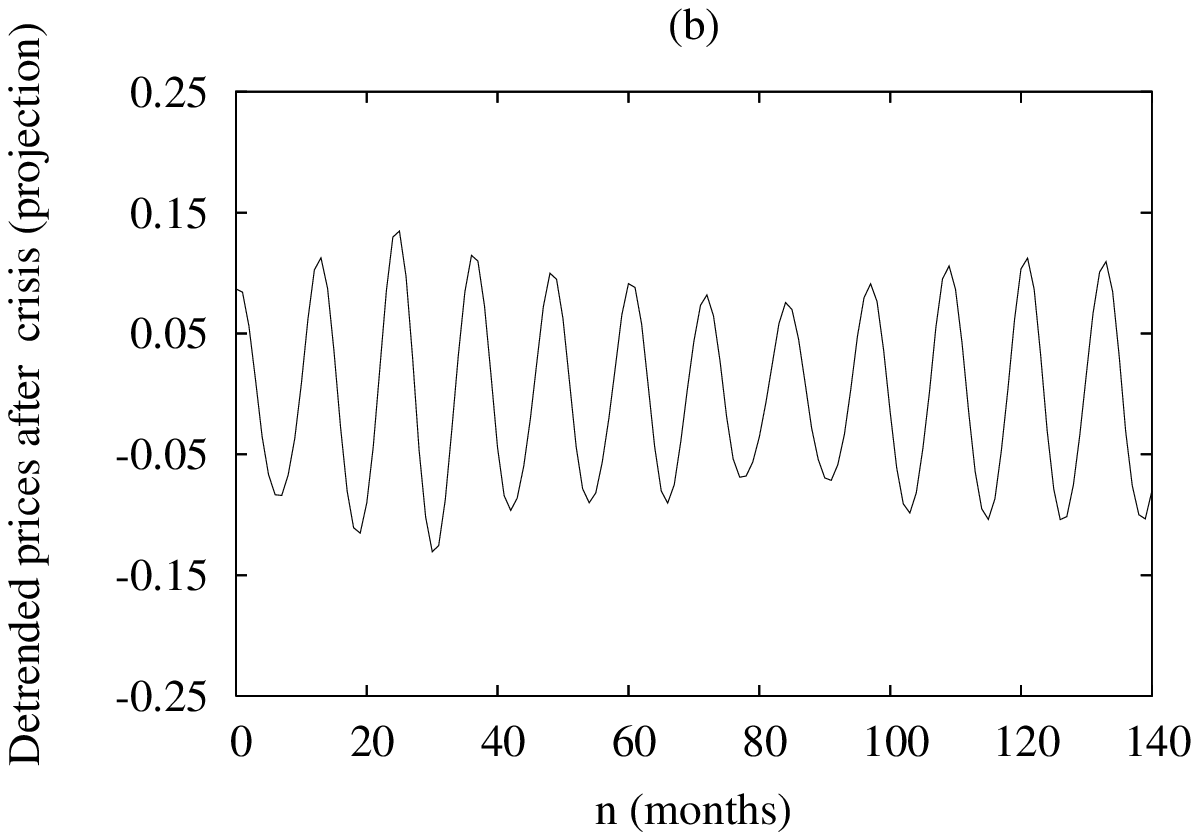}
\end{center}
\begin{center}
\includegraphics[width=7cm]{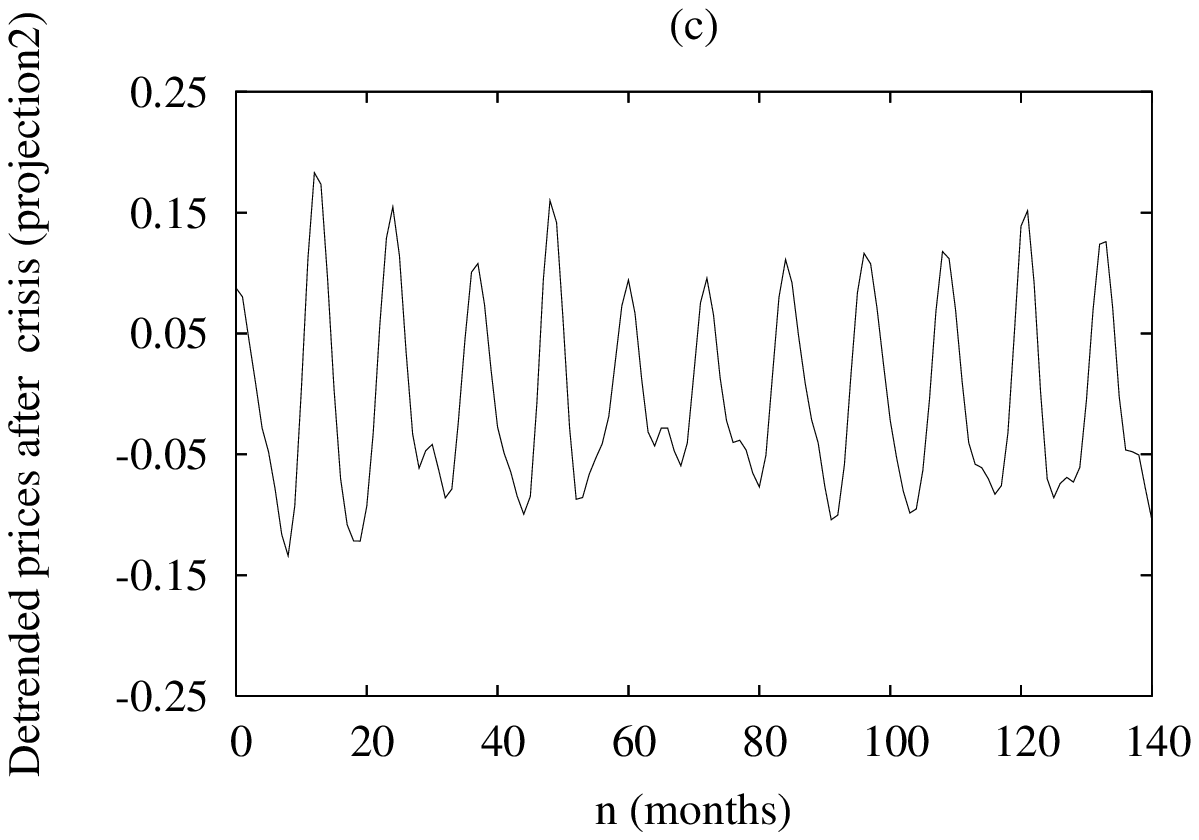}
\end{center}
\vskip1cm
\caption{Time series for the detrended prices after crisis. Panel (a):
full time series. Panel (b): projection of the time series on the first two 
principal components.
Panel (c): projection of the time series on the first five principal components}
\end{figure}
In order to test the last assumption we plot in Fig. 6 the full 
detrended time series after the crisis (panel (a)) as
well as the time series projected in the subspace of first two principal
components (panel (b)) or on the first five principal components (panel (c)).
As we can see essential features of the dynamics such as its periodicity
are captured already in two-dimensional phase space whereas the details of the dynamics of the
fluctuation can be very well described when the $S=5$. On the basis
of this conclusion in future we shall try to model the fluctuation
price dynamics by means of a low-dimensional nonlinear dynamics model.
\par
As concluding remark we note that in this paper we
concentrated our attention on a problem not much investigated
up to now by the econophysics which is focused much more on
the statistical properties of economic and financial systems
containing many similar- (small) size interacting agents. 
Here we show that the presence of a large agent(s) in such
systems can lead to significant changes of the environment of
action and interaction of the smaller agents. We have shown that
as a result of the Japan government intervention on the agriculture markets after 
the oil crisis in 1974 (i) the large pre-crisis nonstationarity of the
piglet prices was reduced; (ii) the price behavior became cyclic and was coupled to the
consumer-driven production cycle and (iii) In addition 
the dimension of the phase 
space of the dynamics of price fluctuations around the yearly moving 
average became smaller as a result of the intervention.
The government intervention has made the prices more 
predictable for the suppliers and for the buyers.  Thus we can
conclude that with respect to this sub-sector of the Japan agriculture the
government intervention was successful. 
\begin{acknowledgments}
N. K. Vitanov thanks for the support of his research by a Fellowship of the Japan 
Society for the Promotion of Science (JSPS) and by the COST P10 "Physics of Risk" Program of the European Science Foundation.
\end{acknowledgments}
\bibliography{vitanov}
\end{document}